\input{aipcheck}

\documentclass[
    ,final            
  ]
  {aipproc}

\layoutstyle{6x9}

\newcommand{\be}{\begin{eqnarray}}
\newcommand{\ee}{\end{eqnarray}}

\begin{document}

\title{Density effects on the pion dispersion relation at finite temperature}

\author{Alejandro Ayala}{
  address={Instituto de Ciencias Nucleares, Universidad
           Nacional Aut\'onoma de M\'exico,  
           Apartado Postal 70-543, M\'exico Distrito Federal 04510,
           M\'exico.}
}

\begin{abstract}
We study the behavior of the pion dispersion relation in a pion medium
at finite density and temperature, introducing a chemical
potential to describe the finite pion number density. Such
description is particularly important during the hadronic phase of a
relativistic heavy-ion collision, between chemical and thermal
freeze-out, where the pion number changing processes, driven by the
strong interaction, can be considered to be frozen. We make
use of an effective Lagrangian that explicitly respects chiral
symmetry through the enforcement of the chiral Ward identities. The pion
dispersion relation is computed through the computation of the pion
self-energy in a non-perturbative fashion by giving an approximate
solution to the Schwinger-Dyson equation for this self-energy. The
dispersion relation is described in terms of a density and temperature
dependent mass and an index of refraction which is also temperature, density 
as well as momentum dependent. The index of refraction is larger than
unity for all values of the momentum for finite $\mu$ and $T$. Given
the strong coupling between $\rho$ vectors and pions, we argue that
the modification of the pion mass due to finite pion density effects
has to be taken into account self-consistently for the description of
the in-medium modifications of $\rho$'s.  
\end{abstract}

\maketitle

 
\section{Introduction}

In recent years, the possibility to produce a locally
thermalized, deconfined state of matter, whose degrees of
freedom are the quarks and gluons of quantum chromodynamics (QCD), the
so called quark-gluon plasma (QGP), in high-energy heavy-ion
collisions, has attracted a great deal of attention, both on the
experimental and the theoretical aspects of the subject~\cite{QM01}.

If the QGP is produced in these kind of reactions, the prevailing view
portrays the evolution of such a system traversing a series of
stages, the last of which consists of a large amount of hadrons
strongly interacting in a finite volume, until a final freeze-out.
For not too high temperatures, hadronic matter consists mainly of
pions, therefore, the study of the propagation properties of pions
within the above described environment represents an important
ingredient for the understanding of the properties of the hadronic
system at and just before freeze-out~\cite{{Shuryak},{Gale-Liu}}. 

The hadronic degrees of freedom are customarily accounted for by means
of effective chiral theories that incorporate the Goldstone boson
nature of pions. One of such theories is chiral perturbation theory
(ChPT) which has been employed to show the well known result that at
leading perturbative order and at low momentum, the modification of
the pion dispersion curve in a pion medium at finite temperature is just
a constant, temperature dependent, increase of the pion
mass~\cite{Gasser}. ChPT has also been used in a two-loop computation
of the pion self-energy~\cite{Schenk} and decay
constant~\cite{Toublan}. A striking result obtained from such
computations is that at second order, the shift in the temperature
dependence of the pion mass is opposite in sign and about three times
larger in magnitude than the first order shift, already at
temperatures close to $150$ MeV. This result might signal either the
breakdown of the perturbative expansion at these temperatures or the need
to perform such calculations using schemes other than perturbation theory.  

However, a missing ingredient in the calculations of the pion
dispersion curve is the treatment of the medium's finite density. The
conceptual difficulty is related to the fact that, though it is
possible to assume that the system is in (at least local) thermal
equilibrium, strictly speaking the only conserved charge that can
be associated to the pion system is the electric charge and thus, for
an electrically neutral pion system the corresponding chemical potential 
vanishes. The behavior of the pion mass in the presence of an isospin
chemical potential has been recently studied in Ref.~\cite{Loewe}. But
in order to describe a situation 
in which the number of pions in thermal equilibrium is finite, we need to
consider a chemical potential associated with the
pion number, instead of its charge. 

Recall that the pion number is not a conserved quantity due to either strong,
weak or electromagnetic processes. Nevertheless, the characteristic
time for electromagnetic and weak pion number-changing processes, is
very large compared to the lifetime of the system created in
relativistic heavy-ion collisions and therefore, these processes are of no
relevance for the propagation properties of pions within the hadronic
phase of the collision. As for the case of strong
processes, it is by now accepted that they drive pion number toward
chemical freeze-out at a temperature considerably higher than the
thermal freeze-out temperature and therefore, that
from chemical to thermal freeze-out, the pion system evolves with the
pion abundance held fixed~\cite{Bebie, Braun-Munzinger}. Under these
circumstances, it is possible to ascribe to the pion density a
chemical potential and consider the pion number as
conserved~\cite{Hung,Chungsik}. In this context, the role of a finite
pion chemical potential into a hadronic equation of state has been recently
investigated in Ref.~\cite{Teaney}.

Furthermore, another important ingredient in the analysis is the well
know fact that in finite temperature field theories with either
massless degrees of freedom or that exhibit spontaneous symmetry
breaking~\cite{Dolan}, the perturbative expansion breaks down and thus
the necessity to implement resummation techniques. 

In this contribution we summarize the effects that the introduction of
a finite pion chemical potential, associated to the finite pion
density, has on the dispersion curve of pions at finite
temperature. Starting from the linear sigma model, we use an effective
Lagrangian~\cite{Ayala} obtained by integrating out the heavy sigma modes and
compute, in a non perturbative fashion, the pion self-energy. We find
that the dispersion curve is modified with respect to the vacuum in a
way described by the introduction of an index of refraction larger
than one, in addition to the thermal and density increase of the pion
mass. Further details can be found in Ref.~\cite{AAA}

\section{Effective Lagrangian}\label{secII}

The Lagrangian for the linear sigma model, including only meson degrees of 
freedom and with an explicit chiral symmetry breaking term, 
can be written as~\cite{Lee}
\be
   {\mathcal{L}}=\frac{1}{2}\left[(\partial{\mathbf{\pi}})^2 +
                (\partial\sigma)^2 - m_\pi^2{\mathbf{\pi}}^2 - 
                m_\sigma^2\sigma^2\right] 
                -\lambda^2 f_\pi\sigma (\sigma^2 + {\mathbf{\pi}}^2) -
                \frac{\lambda^2}{4}(\sigma^2 + {\mathbf{\pi}}^2)^2\, ,
\label{lagrangian}
\ee
where $\mathbf{\pi}$ and  $\sigma$ are the pion and sigma fields,
respectively, and the coupling $\lambda^2$ is
\be
   \lambda^2=\frac{m_\sigma^2-m_\pi^2}{2f_\pi^2}\, .
   \label{coupling}
\ee

When interested in a given approximation scheme to build the Green's
functions of the theory at a given perturbative order, it is possible
to exploit the relations that chiral  
symmetry imposes among them. These relations, better known as chiral
Ward identities (ChWI), are a direct consequence of the fact that the
divergence of the axial current may be used as an interpolating field
for the pion~\cite{Lee}. In order to make sure that the 
approximation respects chiral symmetry, one needs to check that the
modification of other Green's functions respect the corresponding
ChWI. For example, two of the ChWI satisfied --order by order in
perturbation theory-- by the functions  
$\Delta_\pi (P)$, $\Delta_\sigma (Q)$, $\Gamma_{12}^{ij}$ and 
$\Gamma_{04}^{ijkl}$ are
\be
   f_\pi\Gamma_{04}^{ijkl}(;0,P_1,P_2,P_3)&=&
   \Gamma_{12}^{kl}(P_1;P_2,P_3)\delta^{ij} 
   +
   \Gamma_{12}^{lj}(P_2;P_3,P_1)\delta^{ik}
   +
   \Gamma_{12}^{jk}(P_3;P_1,P_2)\delta^{il}\nonumber\\
   f_\pi\Gamma_{12}^{ij}(Q;0,P)&=&
   \left[\Delta_\sigma^{-1}(Q) - 
   \Delta_\pi^{-1}(P)\right]\delta^{ij}\, ,
   \label{Ward}
\ee
where momentum conservation at the vertices is implied, that is
$P_1+P_2+P_3=0$ and $Q+P=0$. The notation for the functional
dependence of the vertices in Eqs.~(\ref{Ward}) is such that the
variables before and after the semicolon refer to the four-momenta of
the sigma and pion fields, respectively~\cite{Lee}.  

In Refs.~\cite{Ayala} it has been shown that in the kinematical regime
where the pion momentum, the pion mass and the temperature are small compared
to the sigma mass, the effective one-loop sigma propagator
and one-sigma two-pion and four-pion vertices are given by 
\be
   i\Delta_\sigma^\star (Q)&=&\frac{i}{Q^2 - m_\sigma^2 + 
                              6\lambda^4f_\pi^2{\mathcal I}^t
                              (Q) }\, .
   \label{newsigprop}
\ee
\be
   i\Gamma_{12}^{\star\, ij}(Q;P_1,P_2)=
   -2i\lambda^2 f_\pi\delta^{ij}
   \left[1 - 3\lambda^2{\mathcal I}^t(Q)\right]\, ,  
   \label{vert12tot}
\ee
\be
   i\Gamma_{04}^{\star\ ijkl}(;P_1,P_2,P_3,P_4)\!\!&=&\!\!
   2i\lambda^2\left\{\right.
   \nonumber\\
   \!\!&\times&\!\!
   \left.\left[1 - 3\lambda^2{\mathcal I}^t(P_1+P_2)\right]
   \delta^{ij}\delta^{kl}\right.\nonumber\\
   \!\!&+&\!\! 
   \left[1 - 3\lambda^2{\mathcal I}^t(P_1+P_3)\right]
   \delta^{ik}\delta^{jl}\nonumber\\
   \!\!&+&\!\! \left.
   \left[1 - 3\lambda^2{\mathcal I}^t(P_1+P_4)\right]\delta^{il}
   \delta^{jk}\right\}\, ,
   \label{vert4tot}
\ee
where in the imaginary-time formalism of thermal field theory (TFT), the 
function ${\mathcal I}^t$ is obtained as the time-ordered analytical
continuation to real energies of the function ${\mathcal I}$ defined by
\be
   {\mathcal I}(Q)\equiv T\sum_n\int\frac{d^3k}{(2\pi)^3}
   \frac{1}{K^2+m_\pi^2}
   \,\frac{1}{(K-Q)^2+m_\pi^2}\, .
   \label{funcIJ}
\ee
Here $Q=(\omega,{\mathbf{q}}),\, K=(\omega_n,{\mathbf{k}})$ are
Eucledian space four-vectors, namely
$Q^2=\omega^2 + q^2$, $K^2=\omega_n^2 + k^2$ with
$\omega = 2m\pi T$ and $\omega_n = 2n\pi T$ ($m$, $n$ integers) being
discrete boson frequencies, $T$ is the temperature and
$q=|{\mathbf{q}}|,\, k=|{\mathbf{k}}|$. 

It is easy to check that Eqs.~(\ref{newsigprop}), (\ref{vert12tot}) and 
(\ref{vert4tot}) satisfy the Ward identities in Eq.~(\ref{Ward}), this ensures
that the approximation scheme adopted respects chiral symmetry.

By using the above effective vertices and propagator, it is possible
to construct the two-loop modification to the pion self-energy in the
same kinematical regime with the result~\cite{Ayala}
\be
   \Pi_2(P)=\left(\frac{m_\pi^2}{2f_\pi^2}\right)
   T\sum_n\int\frac{d^3k}{(2\pi)^3}\frac{1}{K^2+m_\pi^2}
   \left\{3-\left(\frac{m_\pi^2}{2f_\pi^2}\right)
   \left[9{\mathcal I}^t(0) + 6{\mathcal I}^t(P+K)\right]\right\}.
   \label{selfmod}
\ee
Equation~(\ref{selfmod}) reproduces the leading order result obtained from
ChPT~\cite{Gasser}. Furthermore, we observe that Eq.~(\ref{selfmod})
can be formally obtained by means of the effective Lagrangian 
\be
   {\mathcal L}=\frac{1}{2}\left(\partial_\mu{\mathbf{\phi}}\right)^2
   -\frac{1}{2}m_\pi^2{\mathbf{\phi}}^2 -\frac{\alpha}{4!}\left(
   {\mathbf{\phi}}^2\right)^2\, ,
   \label{effLag}
\ee
where $\alpha=6(m_\pi^2/2f_\pi^2)$ and the factor $6$ comes from
considering the interaction of like-isospin pions in the vertex
\be
   i\Gamma_4^{ijkl}=-2i\left(\frac{m_\pi^2}{2f_\pi^2}\right)
   \left(\delta^{ij}\delta^{kl}+\delta^{ik}\delta^{jl}+\delta^{il}\delta^{jk}
   \right)\, .
   \label{vertmod}
\ee
Eqs.~(\ref{selfmod}) and~(\ref{effLag}) mean that in the kinematical
regime where the sigma mass is large compared to the pion mass,
the momentum and the temperature, the linear sigma model Lagrangian
reduces to a $\phi^4$ Lagrangian for effective like-isospin pions with
an effective coupling given by $\alpha$. In essence, the theory thus
constructed and summarized by the effective Lagrangian in
Eq.~(\ref{effLag}) can be thought of a theory for the effective
coupling $\alpha$. By restricting ourselves to
the above kinematical regime, we will proceed on working with the
Lagrangian given by Eq.~(\ref{effLag}). 

\section{Non-perturbative pion self-energy}\label{secIII}

It is well know that quantum field theories at finite temperature
present certain subtleties such as the breakdown of the perturbative
expansion~\cite{Kapusta}. This breakdown becomes manifest in two
important cases: the appearance of infrared divergences in theories
with massless degrees of freedom, and the compensation of powers of the
coupling constant with powers of $T$ for large temperatures. In both
situations, the resummation of certain classes of diagrams represents
an important improvement for the study of the physical properties of such
theories. Even for cases where neither there is a massless degree of
freedom, nor the temperature is extremely large, it is important to
consider a resummation scheme, particularly for the case of theories with
spontaneous symmetry breaking near critical behavior.

In order to consider the above mentioned general situation and with the
purpose of studying the pion dispersion curve at finite
density and temperature, let us formally consider the
Schwinger-Dyson equation for the pion self-energy
\be
   \Pi(P)&=&\frac{\alpha}{2}T\sum_n\int\frac{d^3k}{(2\pi)^3}
   \frac{1}{K^2+m_\pi^2+\Pi}\nonumber\\
   &-&\frac{\alpha^2}{6}T^2\sum_{n_1,n_2}\int
   \frac{d^3k_1}{(2\pi)^3}\frac{d^3k_2}{(2\pi)^3}\frac{1}{K_1^2+m_\pi^2+\Pi}
   \nonumber\\
   &\times&
   \frac{1}{K_2^2+m_\pi^2+\Pi}
   \frac{1}{(K_1+K_2-P)^2+m_\pi^2+\Pi}\, ,
   \label{SD}
\ee
where for internal lines, we make the substitution $i\omega\rightarrow
i\omega +\mu$.
Notice that since the interaction Lagrangian contains only a quartic
term, there is no need to dress the vertices in the above
equation.

Equation~(\ref{SD}) represents an integral equation for the function
$\Pi(P)$, which, needless to say, is very difficult to be solved
exactly. In order to find an approximate solution let us write 
\be
   \Pi(P)\simeq\Pi_0+\tilde{\Pi}(P)
   \label{decomp}
\ee
and consider $\tilde{\Pi}(P)\ll\Pi_0$. As we will see, such assumption
is justified given that in our approximation, $\tilde{\Pi}\sim{\mathcal
O}(\alpha^2)$. Keeping only the lowest order contribution in
$\tilde{\Pi}$, we find
\be
   \Pi_0\equiv\frac{\alpha}{2}T\sum_n\int\frac{d^3k}{(2\pi)^3}
   \frac{1}{K^2+m_\pi^2+\Pi_0}\, ,
   \label{pi0}
\ee
\be
   \tilde{\Pi}(P)&\equiv&-\frac{\alpha^2}{6}T^2\sum_{n_1,n_2}\int
   \frac{d^3k_1}{(2\pi)^3}\frac{d^3k_2}{(2\pi)^3}\frac{1}{K_1^2+m_\pi^2+\Pi_0}
   \nonumber\\
   &\times&
   \frac{1}{K_2^2+m_\pi^2+\Pi_0}
   \frac{1}{(K_1+K_2-P)^2+m_\pi^2+\Pi_0}\, .\nonumber\\
   \label{pitilde}
\ee
Equation~(\ref{pi0}) represents a self consistent equation for the
(momentum independent) constant $\Pi_0$. This is the well known
resummation for the {\it superdaisy} diagrams which constitute the dominant
contribution in the large-$N$ expansion~\cite{Dolan} of the Lagrangian in
Eq.~(\ref{effLag}). On the other hand, Eq.~(\ref{pitilde}) represents a first
approximation to the momentum dependent piece of $\Pi (P)$. 

\begin{figure} 
\begin{tabular}{cc}
\includegraphics[height=.22\textheight]{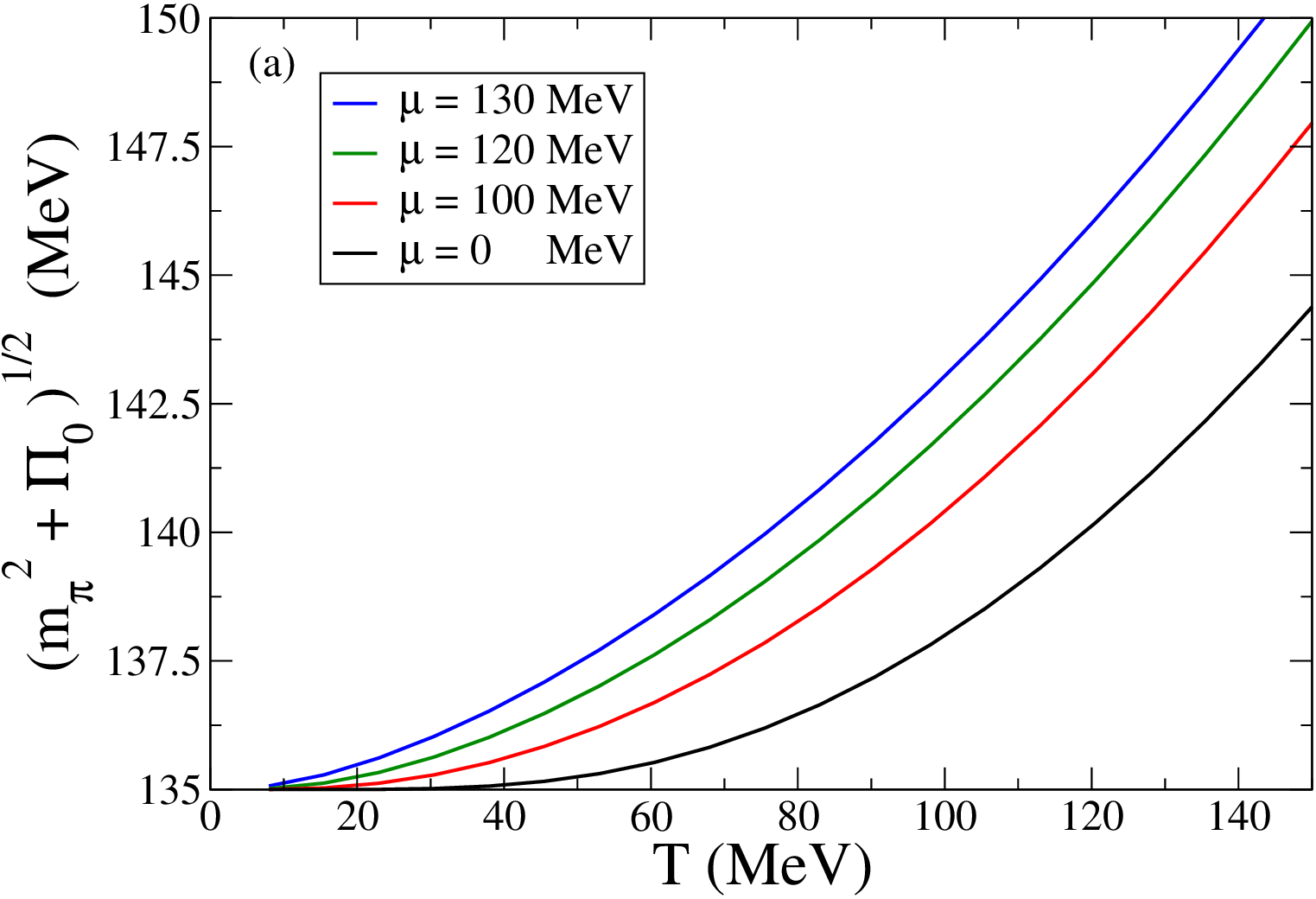}
&
\includegraphics[height=.22\textheight]{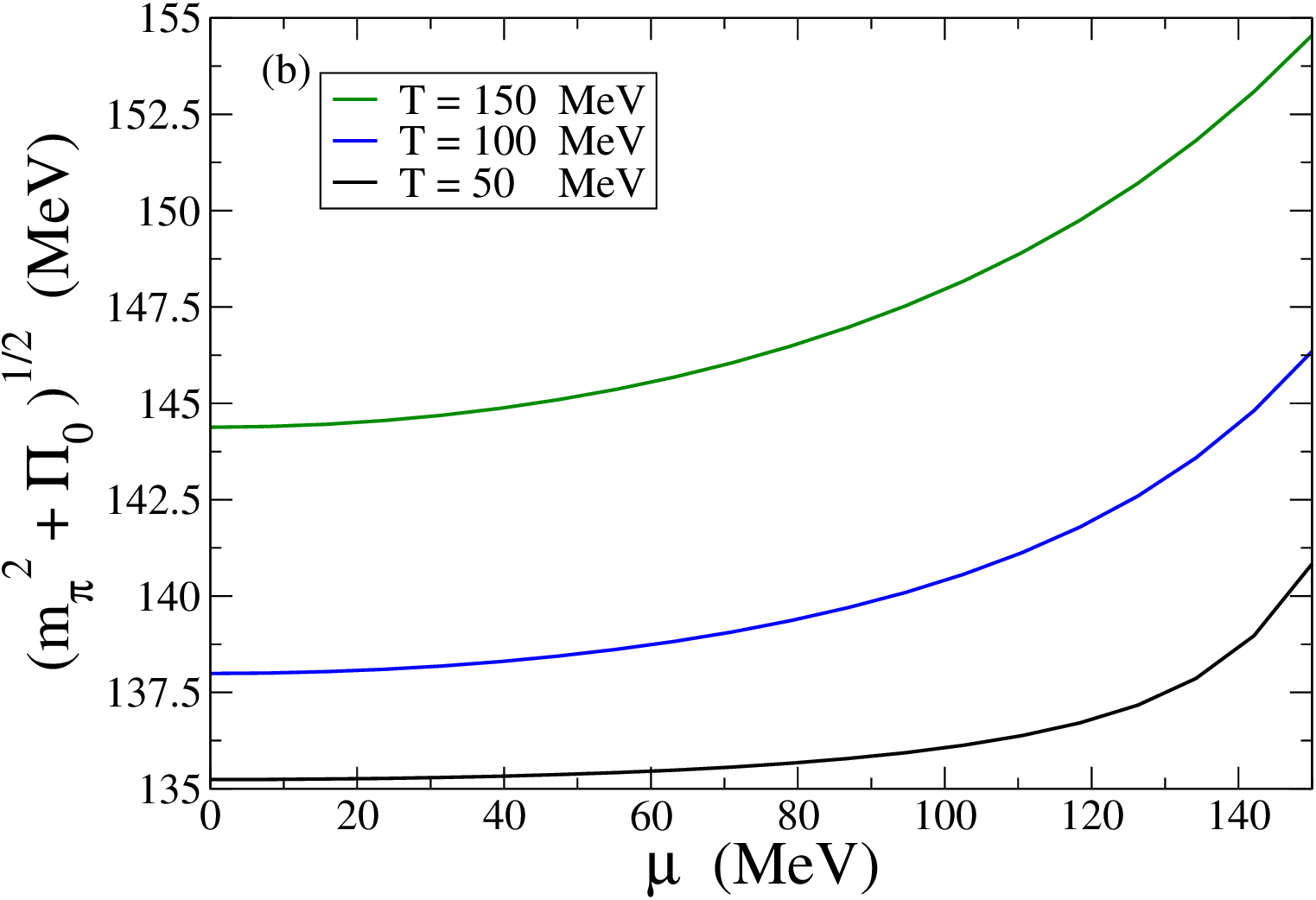} \\
\end{tabular}
\caption{$\sqrt{m_\pi^2 + \Pi_0}$ as a function of $(a)$ the
temperature $T$ for different values of the chemical potential ranging
from $\mu=0$ to $\mu=130$ MeV, from bottom to top, and as a function of
$(b)$ the chemical potential $\mu$ for different values of the
temperature ranging from $T=50$ to $T=150$ MeV, from bottom to top.}
\end{figure}

Figure~1 shows the behavior of the quantity $\sqrt{m_\pi^2 + \Pi_0}$
as a function of $(a)$ the temperature $T$ for different values of the
chemical potential $\mu$ and $(b)$ as a function of $\mu$ for
different values of $T$. In both cases, $\sqrt{m_\pi^2 + \Pi_0}$ grows
monotonically with both $T$ and $\mu$.

The pion dispersion relation is given as the solution to
\be
   p_0^2-\left[p^2+m_\pi^2+\Pi_0 +
   {\mbox R}{\mbox e}\tilde{\Pi}^r(p_0,p)\right]=0
   \label{solution}
\ee
for positive $p_0$. In Eq.(\ref{solution}), $\tilde{\Pi}^r$ denotes
the retarded version of $\tilde{\Pi}$ upon analytical continuation to
Minkowski space. The left panel of Fig.~2 shows plots of $p_0$ as a
function of $p$ for different values of $\mu$ and a temperature
$T=120$ MeV. As can be seen from this figure, $\Pi(p_0,p=0)$
contributes to the increase of the pion mass. Also, for large $p$, the
dispersion curves approach the light cone, always from within the
causal region $p_0^2 > p^2$. 

\begin{figure} 
\begin{tabular}{cc}
\includegraphics[height=.22\textheight]{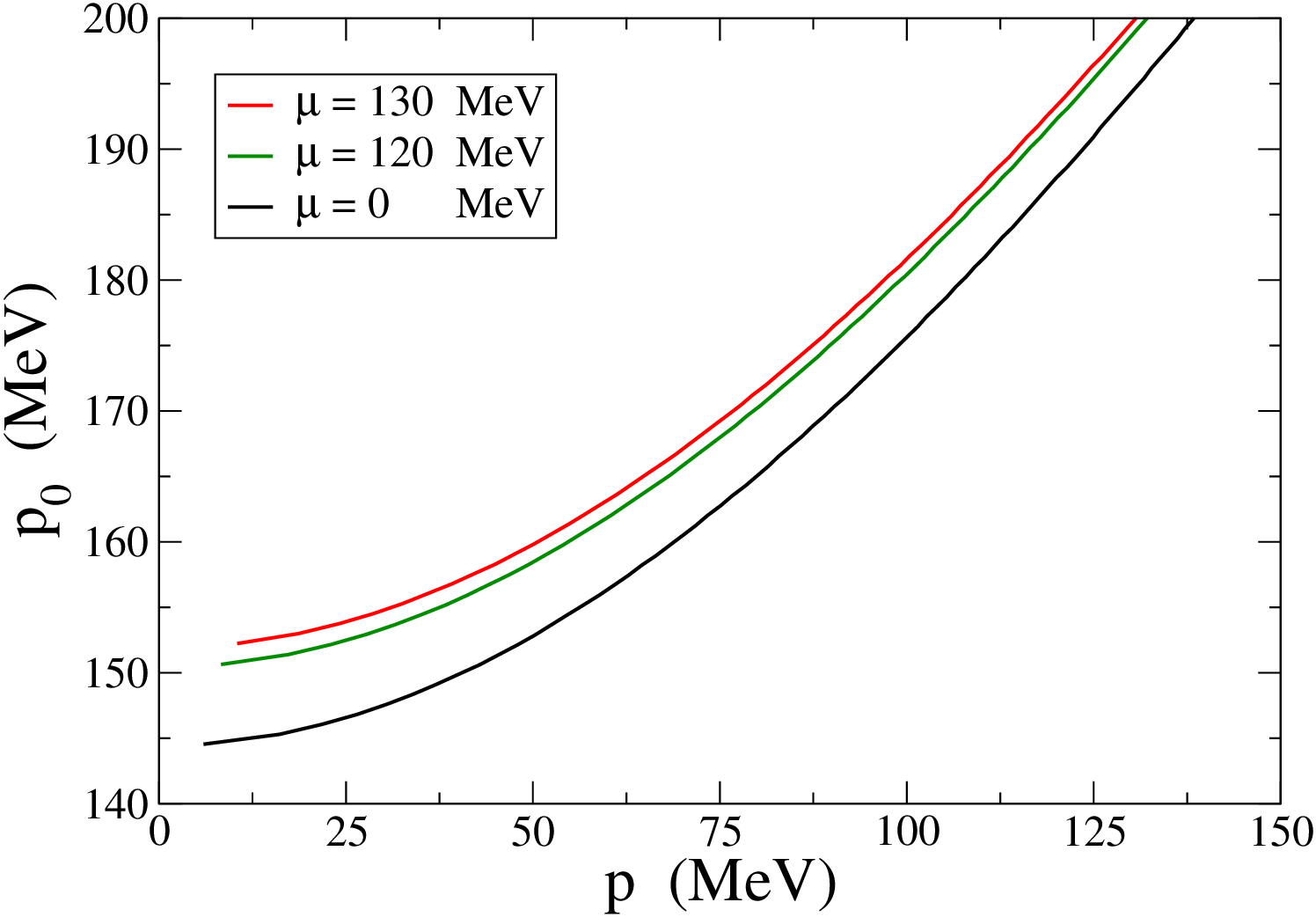}
&
\includegraphics[height=.22\textheight]{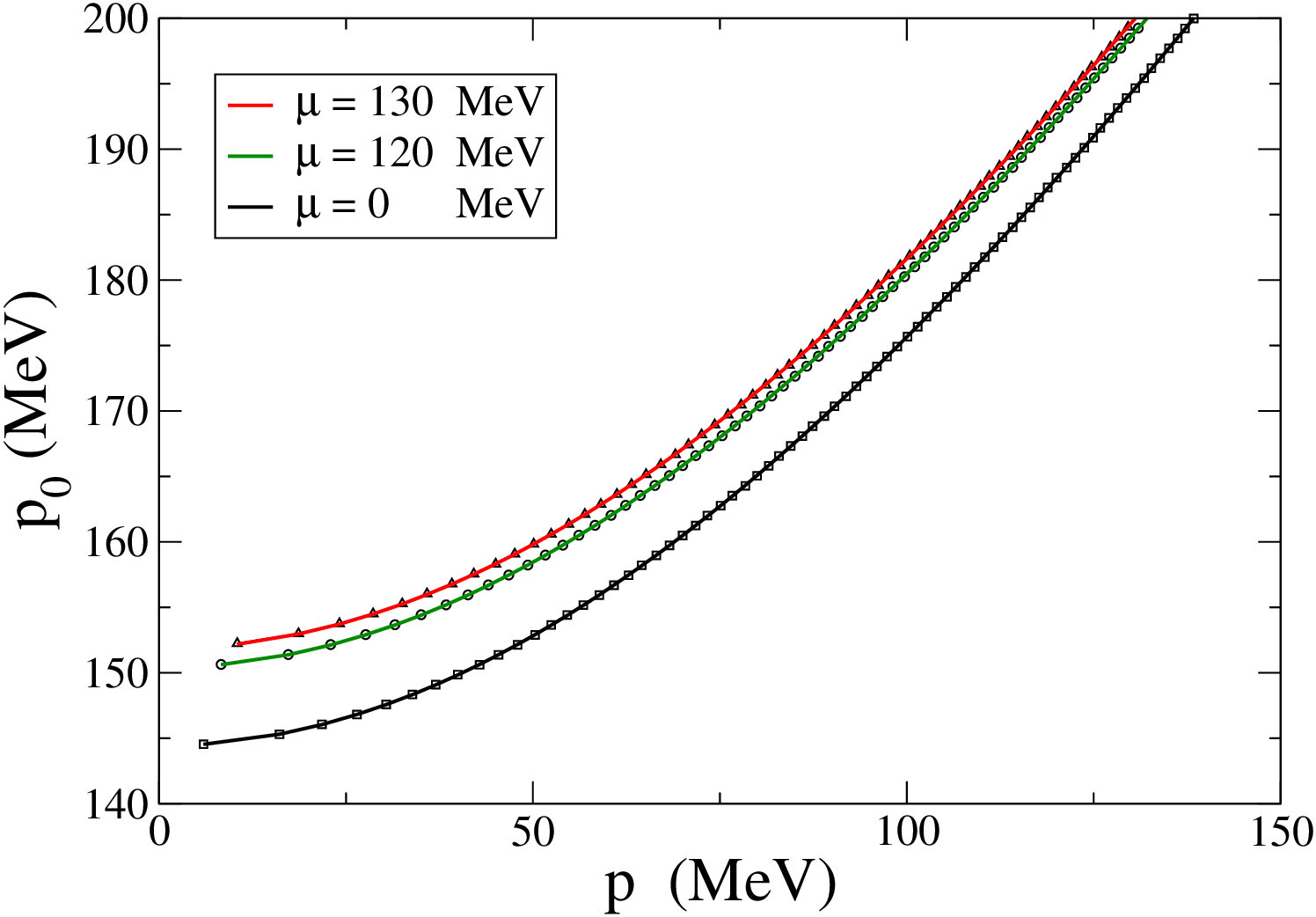} \\
\end{tabular}
\caption{Left panel: $p_0$ as a function of $p$ for $\mu=0,120,130$
MeV, from bottom to top, and $T=120$ MeV. Right panel: Fit of the
dispersion curve to the functional form $p_0=\sqrt{n^{-1}(T,\mu)p^2 +
M^2(T,\mu)}$ for $\mu=0,120,130$ MeV, from bottom to top, and $T=120$
MeV. The dots are the computed values of the dispersion relation.}
\end{figure}

We also notice that the dispersion curves can be parametrized by a
function of the form
\be
   p_0=\sqrt{n^{-1}(T,\mu)p^2 + M^2(T,\mu)}\, .
   \label{parametri}
\ee
This is shown in the right panel of Fig.~2 where the dots represent
the computed values of the dispersion relation and the continuous
curves the fits. We thus observe that the presence of a finite
chemical potential has two dramatic effects on the behavior of the pion
dispersion curve compared to the case where only the temperature is
considered: first, there is a significant increase in the pion mass
and second, the index of refraction parameter $n$ becomes larger than unity.

\section{Summary and conclusions}\label{secIV}

In this paper we have considered the effects of a finite pion density
on the pion dispersion curve at finite temperature. The finite density
is described in terms of a finite pion chemical potential. We have argued
that such description is important during the hadronic phase of a
collision of heavy nuclei at high energies between chemical and
thermal freeze-out when the strong pion-number changing processes have
driven the pion number to a fixed value. 

In order to consider a general scenario that takes into account
resummation effects, we have presented an approximate solution to the
Schwinger-Dyson equation for the momentum-dependent pion
self-energy, writing this as $\Pi(P)\simeq\Pi_0+\tilde{\Pi}(P)$ and
considering $\tilde{\Pi}(P)\ll\Pi_0$, which 
is justified given that in our approximation, $\tilde{\Pi}\sim{\mathcal
O}(\alpha^2)$, whereas the perturbative expansion of $\Pi_0$ starts at
${\mathcal O}(\alpha)$.
 
The pion dispersion relation thus obtained at finite density and
temperature deviates from the vacuum dispersion relation and can be
described in terms of a density and temperature dependent mass and an
index of refraction, also temperature, density as well as momentum
dependent. This index of refraction is larger than unity for all
values of the momentum. Similar results have been obtained in
Ref.~\cite{Pisarski} using also a linear sigma model {\it without} the
introduction of a finite chemical potential but rather as a
consequence of the loss of Lorentz invariance when a particle travels in a
medium at finite temperature. We note however that our result is more
general than that of the former reference where the computation of the
pion dispersion relation was carried from the onset in the weak coupling
regime, that is to say, $\lambda^2 \ll 1$, whereas our approach is
valid for arbitrary values of $\lambda^2$~\cite{Ayala}.

Recall that one of the most salient features resulting from the
analyses of collisions of heavy nuclei at SPS energies is the low
mass dilepton spectra which shows an increase around the free $\rho$
peak. Such behavior has been explained in terms of possible in-medium
modifications of the $\rho$ with either a shift toward lower values
of its mass or an increase in its width~\cite{Rapp}. Given that the
$\rho$ meson couples strongly to the two pion channel, any possible
in-medium modifications of the pion propagation properties should
translate into induced modifications of the $\rho$ meson properties
and should therefore be properly accounted for in a self-consistent manner
in these kind of analyses. In particular, the increase of the pion
mass with density modifies the phase space for the production and
decay of $\rho$'s in equilibrium and consequently it affects the
production of $e^+\ e^-$ pairs. This matters will be the subject of a
future work~\cite{progress}.  

\section*{Acknowledgments}

Support for this work has been received in part by DGAPA-UNAM under
PAPIIT grant number IN108001 and CONACyT under grant numbers 35792-E
and 32279-E.

\vfill\eject
\end{document}